\documentclass[sigconf]{acmart}
\renewcommand\footnotetextcopyrightpermission[1]{}

\usepackage{booktabs} 

 \setcopyright{none}

\acmDOI{10.475/123_4}

\acmISBN{123-4567-24-567/08/06}

\usepackage{amsmath,amssymb,amsfonts} 
\usepackage{algorithmic} 
\usepackage{graphicx} 
\usepackage{textcomp} 
\usepackage{xcolor} 
\usepackage{amssymb} 
\usepackage[many]{tcolorbox} 
\usepackage{mathtools} 
\usepackage{tabularx} 
\usepackage{booktabs} 
\usepackage{subcaption} 
\usepackage{listings} 
\usepackage{longtable} 
\usepackage{multirow} 
\usepackage{multicol} 
\usepackage{float}
\usepackage{url} 
\usepackage{enumitem}
\newtcolorbox{mybox}[1]{%
    tikznode boxed title, 
    enhanced, 
    arc=0mm, 
    interior style={white}, 
    attach boxed title to top center= {yshift=-\tcboxedtitleheight/2}, 
    fonttitle=\bfseries, 
    colbacktitle=white, coltitle=black, 
    boxed title style={size=normal,colframe=white,boxrule=0pt}, 
    title={#1}}
    
\begin{document}
\title{What is my data worth? From Data Properties To Data Value}

\author{Kalapriya Kannan}
\affiliation{%
  \institution{IBM Research, AI}
  \city{Bengaluru}
  \country{India}
}
\email{kalapriya.kannan@in.ibm.com}

\author{Rema Ananthanarayanan}
\affiliation{%
  \institution{IBM Research, AI}
  \city{Delhi}
  \country{India}
}
\email{arema@in.ibm.com}

\author{Sameep Mehta}
\affiliation{%
  \institution{IBM Research, AI}
  \city{Bengaluru}
  \country{India}
}
\email{sameepmehta@in.ibm.com}

\begin{abstract}
Data today fuels both the economy and advances in machine learning and AI. All aspects of decision making, at the personal and enterprise level and in governments are increasingly data-driven. Vast quantities of data are spurring the development of various analytic techniques in machine learning, which in turn create the demand for more data for model building and validation. In this context, however, there are still some fundamental questions that remain unanswered with respect to data. \textit{What is meant by data value? How can it be quantified, in a general sense?}. The ``value'' of data is not understood quantitatively until it is used in an application
and output is evaluated, and hence currently it is not possible to assess the value of large amounts of data that companies hold, categorically. Further, there is overall consensus that good data is important for any analysis but there is no independent definition of what constitutes good data.

In our paper we try to address these gaps in the valuation of data and present a framework for users who wish to assess the value of data in a categorical manner. Our approach is to view the data as composed of various attributes or characteristics, which we refer to as facets, and which in turn comprise many sub-facets. We define the notion of values that each sub-facet may take, and provide a seed scoring mechanism for the different values. The person assessing the data is required to fill in the values of the various sub-facets that are relevant for the data set under consideration, through a questionnaire that attempts to list them exhaustively. Based on the scores assigned for each set of values, the data set can now be quantified in terms of its properties. This provides a basis for the comparison of the relative merits of two or more data sets in a structured manner, independent of context. The presence of context adds additional information that improves the quantification of the data value. 

\end{abstract}

%
%


\maketitle

\section{Introduction}
\label{intro}

\begin{quote}
 \emph {`` I often say that when you can measure what you are speaking about, and express it in numbers, you know something about it; but when you cannot measure it, when you cannot express it in numbers, your knowledge is of a meagre and unsatisfactory kind. "\footnote{Lord Kelvin, in PLA, vol. 1, "Electrical Units of Measurement", 1883-05-03}}
 \end{quote}
The expanding areas of machine learning increasingly depend on the availability of different kinds of data sets for building and training new advanced models. The right data is a critical component at all stages of decision making, and is a crucial part of the process pipeline for descriptive, predictive and prescriptive analytics. As decision makers are increasingly attempting to better understand the value of the data they hold, it raises some fundamental questions on how the data is to be evaluated. The classical approach has been, and still is, to evaluate the value of data in the context of an application, in terms of how well the contents and quality of the data meet the requirements of the application under discussion. In the non-digital era this was a practical approach, given the number of data sets one had access to, the processing speeds and expectations from business intelligence. Today. companies are evaluated on billions of dollars on the strength of the data they possess and its potential to turn into profit\footnote{The purchase of LinkedIn by MicroSoft, for \$26billion, in 2016 is a case in point.}. Stringent regulations like General Data Protection Regulations necessitate that companies understand the value of the data they hold, and the risks it may pose. Increasing number of data sets are available in the open source domain, and people building data models may have a deluge of data sets to choose from.  In this context, it is important that we be able to answer certain basic questions like \emph{`How valuable is the data I hold?'}, \emph{`I need this data and access to it is available at a cost $C$. Is it worth the cost?'} or, in general, \emph{`Of these multiple data sets, which ones are relatively more valuable?'}. Traditionally, these have all been attempted in the specific context of a set of applications. However we argue here that given the increasing importance of data on all fronts, we need a categorical way of evaluating data, independent of the end application. When additional context is available, we may be able to refine the computed value, but in the absence of this, we should still be able to answer some of the above questions, wholly or partially. This is the goal of our paper and we now introduce the work in detail. in the following set of steps.
\begin{enumerate}
\item We define an exhaustive set of facets for a data set, where a facet is an aspect or attribute of the data set that a user may be interested in, and where the owner or creator of the data set will be able to answer with certainty the value for the facet. \footnote{By 'exhaustive' we intend to cover as many of the aspects of a data set that are common across a set of typical analytic applications. We hope that this will constitute an initial list which can then be extended with further details by various practitioners.}
\item For each facet, we define a set of sub-facets, and for each sub-facet we enumerate the range of possible values that the data set could have for that aspect, and also define an ordering for the values for each of the aspects.
\item Using the relative weights, we describe how to compare the relative merits of two data sets.
\end{enumerate}

There has been some aspects of work that has looked at the creation of a datasheet whenever a new data set is published~\cite{datasheets}. As per this, every data set would be accompanied by a data sheet that documents why it was created, by whom, the composition, the intended uses, maintenance aspects and other properties. The specific goal stated is to improve transparency and accountability in the machine learning community and in some cases to improve the overall quality. Our work has some overlapping areas, but the context, scope and details differ significantly. Our goal is to provide a tool or mechanism that will help users evaluate data sets in general, in terms of the value they could derive from the different data sets. Further, the scope of our work is not limited to openly published data or data of interest curated specifically for various machine learning models, but to data from any source.
To the best of our  knowledge we are not aware of any other similar work that attempts to bootstrap a data set with a value, independent of the end use.
Our main contributions in this paper are as follows.
\begin{enumerate}
    \item We define various attributes of interest across different data sets in general, and typical ranges of values for these attributes. We also define an ordering of the values for each of the attributes.
    \item We then give a system of ranking the relative importance of various attributes (across a common set of applications).
    \item Using this system, we define a means of comparison across two or more data sets, when the facets are listed and values populated
\end{enumerate}

In absence of prior work,  by proposing a method to describe the data set value in an application-agnostic manner, we hope to set the stage for further work in this area.
Given the lack of rigorous definitions of some of the terms and the apparent tight coupling between the data characteristics and the end-application, it is understandable why  much progress has not been made on this front so far. However given the importance of data in all fields of economic activity and governance today, it is essential to have some means of computing the value of a data set independent of the application. The steps described in  the subsequent sections are an initial attempt in this direction. In Section~\ref{background} we present the background and also describe the terms and definitions that we use in the rest of the paper. In Section~\ref{observations} we discuss our observations on the distribution of some metadata values across data sets for a couple of data repositories. In Section~\ref{system} we present the high-level view and then discuss the various properties in detail, and the details of our approach. In Section~\ref{assessment} we discuss some scenarios with examples, of assessing data value. In Section~\ref{relatedwork} we discuss the related work and we conclude in Section~\ref{conclusion}.

\section{Background}
\label{background}

\begin{table*}[!htbp]
\begin{scriptsize}
    \begin{tabular}{|l|p{0.8\textwidth}|}
    \hline
    Facet & Brief description \\
    \hline
     Composition & This is a breakup of the data types of the different fields that comprise one instance.\\
    \hline
    Current usage & This is a statement of whether the data already has known uses.\\
    \hline
    Data quality & This refers to various aspects of data quality such as accuracy, completeness, presence of noise, presence of duplicates and others.\\
    \hline
     Data age & This reflects on the recency or currency of the data. Some data, by nature of its definition,  gets out-dated faster than other data.\\
    \hline
   
    Data sensitivity & This reflects whether the data has confidential information, information relating to health, personal identifiable information and related features that limits its possibility of being shared and restricts the kinds of applications that can be run on it. Such data shares the risk of potential breaches, similar to data that has legal implications.\\
    \hline
     Ease of use & This may be comprised of multiple factors, that overlap with format, size and availability of tools to process the data, for instance. \\ 
    \hline
     Enterprise aspects &  These are typically those aspects of data that an enterprise may use to assess potential use, based on typical use-cases.\\ 
    \hline
      Format & This captures aspects such as the file type, presence of schema and related properties.\\
    
    \hline
 
    Granularity & This is the level of detail provided by instances of the data set and by the data set as a whole.\\
    \hline
    Legal aspects & There may be various legal restrictions on using the data in the current format, or on the audience that may use the data. Such data also poses a higher risk in case of potential breaches and the holder of the data  needs to factor in the costs of higher risk.\\
    \hline
    Purpose & This refers to the reason why the data was collected, obtained or generated.\\
    \hline
     Sourcing & Information on the source could also throw light on potential alternatives available, structure, and other information.\\
    \hline
   
    Statistical properties & This refers to the utility of data for various kinds of statistical analysis.\\
    \hline
    Structure & This states the format of the contents of the data set. Typically this includes structured data, unstructured data and semi-structured data. While structured and semi-structured text lend themselves to querying and a host of other operations because of the wide availability of tools, unstructured text may contain insights not captured in the structured text, and may require pre-processing before it can be used.\\
    \hline
      Tooling & The availability of tools to handle various aspects of processing, querying and maintenance adds to the value of the data.\\
    \hline
  
     Transformations & This refers to any conversions the data may have undergone from one format to another. It is a measure of how much the data has probably been cleansed or brought closer to a form where it can be used for an end-application.\\
    \hline
   
    Uniformity  & This refers to uniformity in structure, where multiple instances that make up  the data set could be similar or dissimilar.\\
    \hline
    Variety & This refers to the number of types of data. \\
    \hline
    Velocity & This is a measure of the rate at which the data arrives. It influences the design of the processing systems that need to handle the data and the storage among other things.\\
    \hline
    Volume & This refers to the size of the data.\\
    \hline
    
    \end{tabular}
    \caption{Facets of data sets\label{tab:facets}}
    \vspace{-0.3in}
    \end{scriptsize}
\end{table*}
In this section, we will discuss the various aspects of a data set that impinge on data value in general, and lay the framework for our subsequent work.

\subsection{Data as an intangible asset}

Data is an intangible asset.  It acquires value only when it is put to use. Unlike other assets, data has only an initial creation cost; once created, there is at most only marginal cost in using it in other applications. In many cases it is created as an intermediate step of some other business process.  Data has a processing cost, in terms of the effort involved in transforming the raw data into the format required for the application. The value of data increases as it moves in this processing pipeline and the same data can be used to prime multiple applications, at various stages. 


\subsection{Aspects of a data set}
Data has been characterized in multiple ways, largely influenced by the purpose for which it is being used. At the simplest level, is the physical format in which it is stored on the various storage devices and made available for higher level programs. The growth and adoption of relational databases led to the characterization of data as structured, unstructured and semi-structured data. The growth of the internet led to the development of the various MIME types (Multipurpose Internet Mail Extensions) which are standardized ways to represent the nature and format of the document, and hence an expression of the data in the document~\cite{mimetypes}. These aspects of the data are intrinsic to the data set, as are other aspects such as the size of the data and certain aspects of data quality that are defined independent of the application that is going to use the data. The 3 V's of big data, volume, variety and velocity, are other characterizations. The classification of data based on the type of content, such as twitter data, online reviews, server logs, Java code or python scripts for instance, is another characterization of the data. For any specific data set, the owner of data will be able to state with preciseness certain aspects of the data, such as the format or size, and the domain it pertains to, independent of the intended end use.

We draw the distinction between intrinsic characteristics or those that are essential or inherent to the data and impartial to the application, and extrinsic characteristics which may be simply defined as those that are not intrinsic. In the context of data
an intrinsic property is a property that can be assessed to be present by the holder of the data, independent of any knowledge of the end-use to which it may be put; properties such as size of the data and its format are intrinsic. To the end-user there may be some intrinsic properties that he cares about and some that he does not care about.  The extrinsic aspects of the data set include certain aspects of data quality that are determined purely by the end-user's application. For instance, the currency of the data,  in terms of how recent it is, is one example that will be defined based on the proposed use. Our hypothesis is that there will be some values of both intrinsic and extrinsic properties that are more valuable in general, even though there are special cases that defy the norm. We will use these values as our heuristic for what represents a good data set, and present an initial evaluation based on this. As more information comes in, it is possible to fine-tune the heuristics in line with the requirements of the actual application. Further, we have been talking of data so far, whereas the logical unit of interest is a data set. We define below the following terms.
\begin{itemize}

    \item{Data set:} A data set is a collection of data. Commonly it refers to a set of related data, such as the contents of a database table or a spreadsheet, or the data in one logical file. It is also used more fluidly to refer to a set of database tables for a specific application, or a set of csv files related to some topic.  A data set is made up of one or more instances of data.
    \item{Data instance:} A data instance is a specific instance of a data item in a data set. It could be an n-tuple, or a single field. A data set has one or more instances. Further. all instances may be of the same type or different types. Each instance may have one or more data points.
\end{itemize}
We use the term \emph{facet} to refer to a dimension of the data that defines some aspect of the data. Each facet could further be divided into sub-facets. In Table~\ref{tab:facets} we list the various facets of data sets, and a high-level description of the same. The sub-facets are enumerated along with possible values, in the next section.

\section{Observation from Public Data Sets}
\label{observations}


\begin{figure*}[ht!]
\centering
  \begin{subfigure}[b]{0.29\textwidth}
    \includegraphics[scale=0.25]{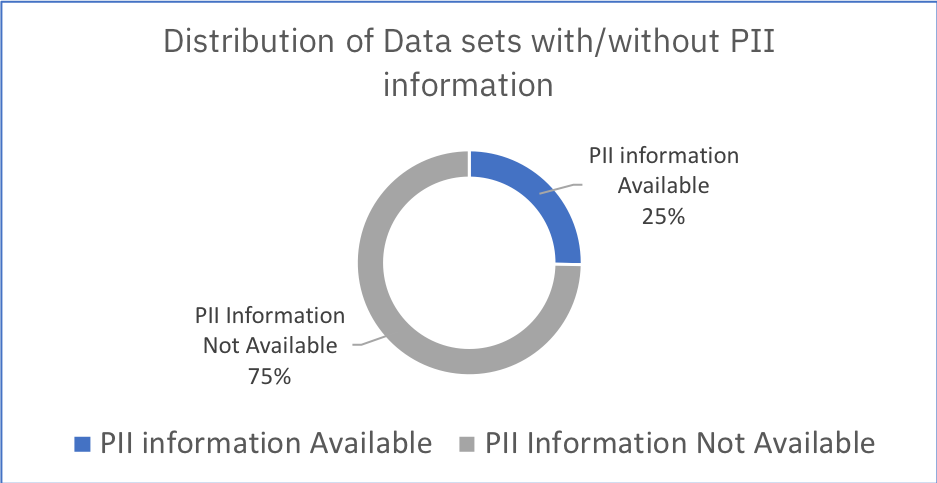}
    \caption{PII information Distribution}
    \label{fig:pii}
  \end{subfigure}
  \begin{subfigure}[b]{0.29\textwidth}
    \includegraphics[scale=0.25]{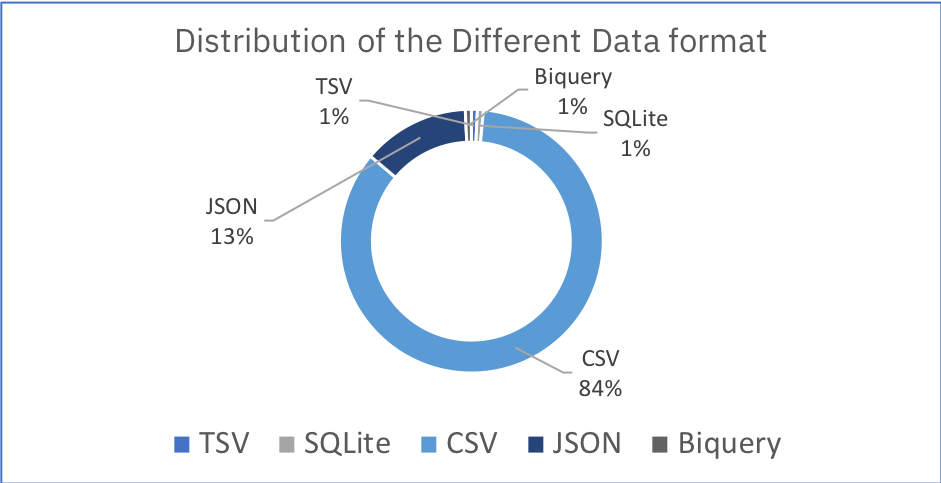}
    \caption{Format Distribution}
    \label{fig:format}
  \end{subfigure}
\begin{subfigure}[b]{0.29\textwidth}
    \includegraphics[scale=0.25]{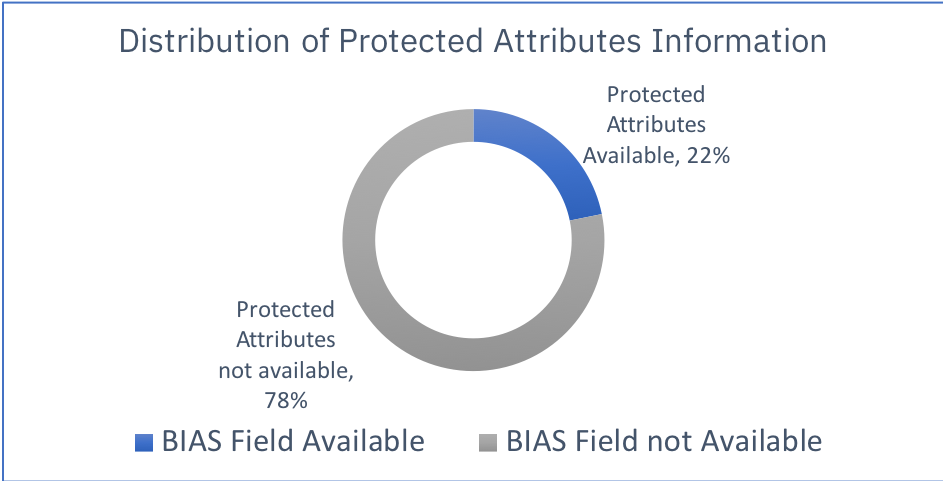}
    \caption{Protected Attributes Distribution}
    \label{fig:bias}
  \end{subfigure}
  \\
  \begin{subfigure}[b]{0.29\textwidth}
    \includegraphics[scale=0.25]{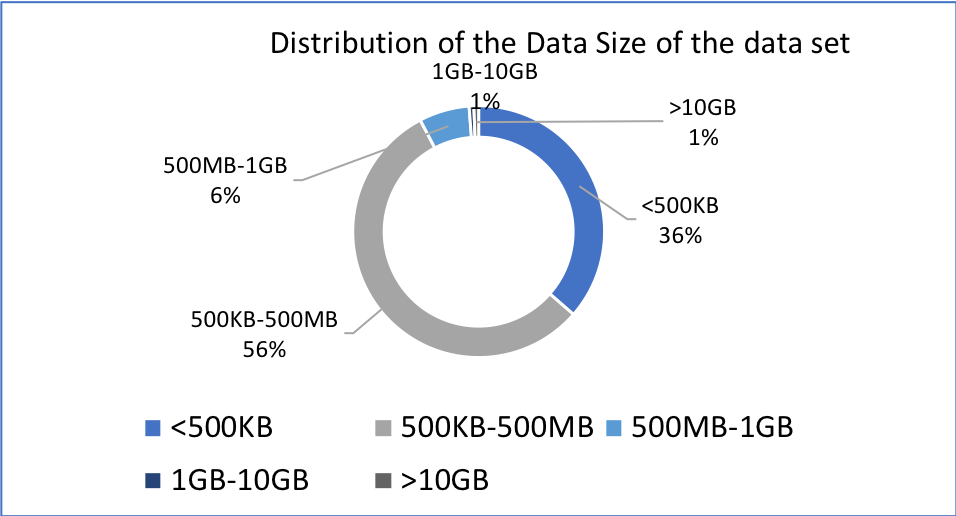}
    \caption{Data Size}
    \label{fig:datasize}
  \end{subfigure}
   \begin{subfigure}[b]{0.29\textwidth}
    \includegraphics[scale=0.25]{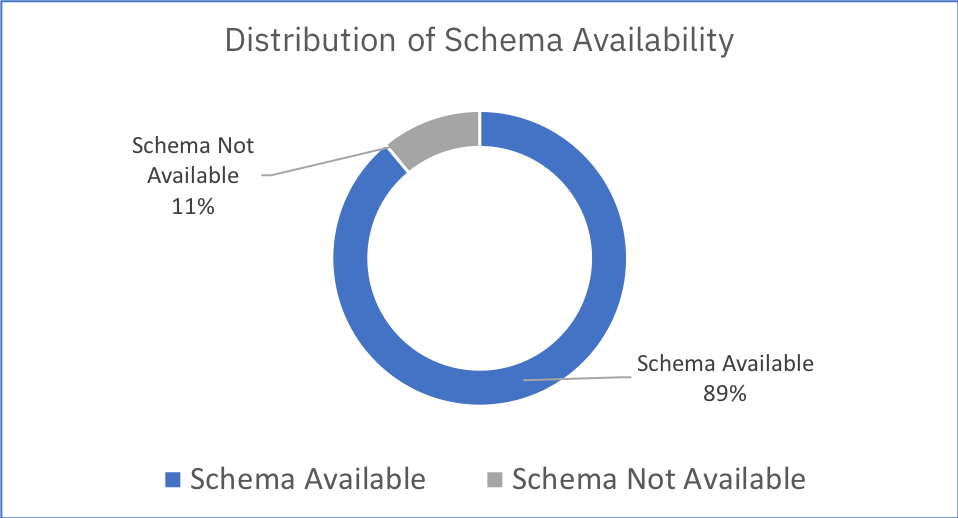}
    \caption{Schema Availability}
    \label{fig:schema}
  \end{subfigure}
\begin{subfigure}[b]{0.29\textwidth}
    \includegraphics[scale=0.25]{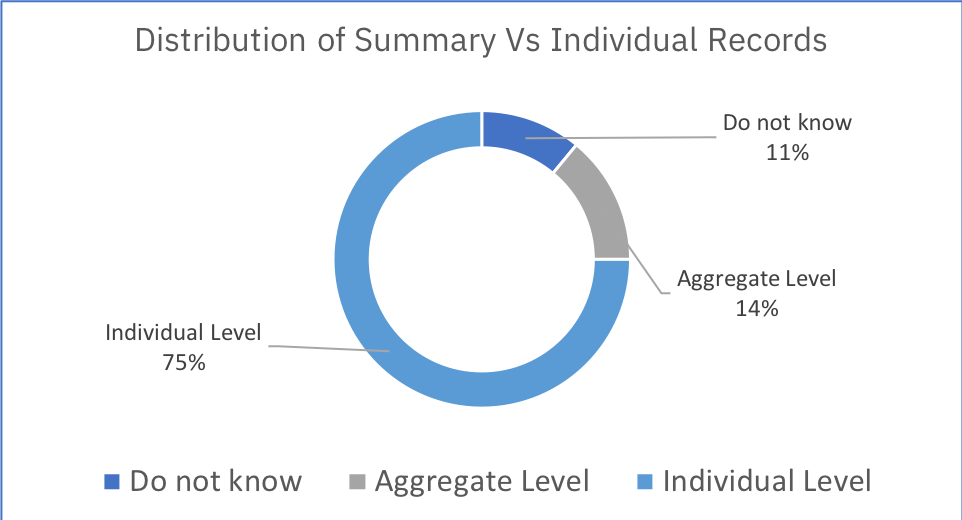}
    \caption{Data Level - Summary/Individual }
    \label{fig:summary}
  \end{subfigure}
  \\
  \begin{subfigure}[b]{0.29\textwidth}
    \includegraphics[scale=0.25]{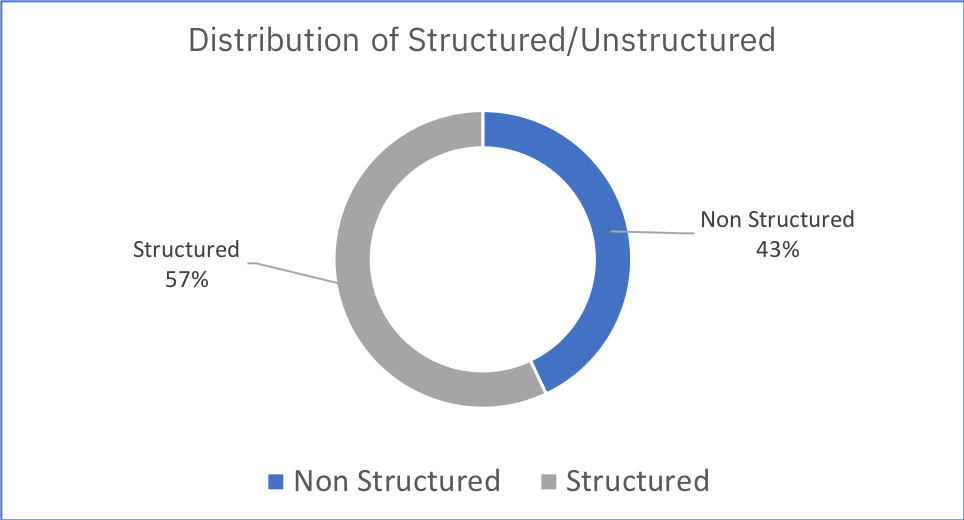}
    \caption{Structured Vs Unstructured}
    \label{fig:summary}
  \end{subfigure}
   \begin{subfigure}[b]{0.29\textwidth}
    \includegraphics[scale=0.25]{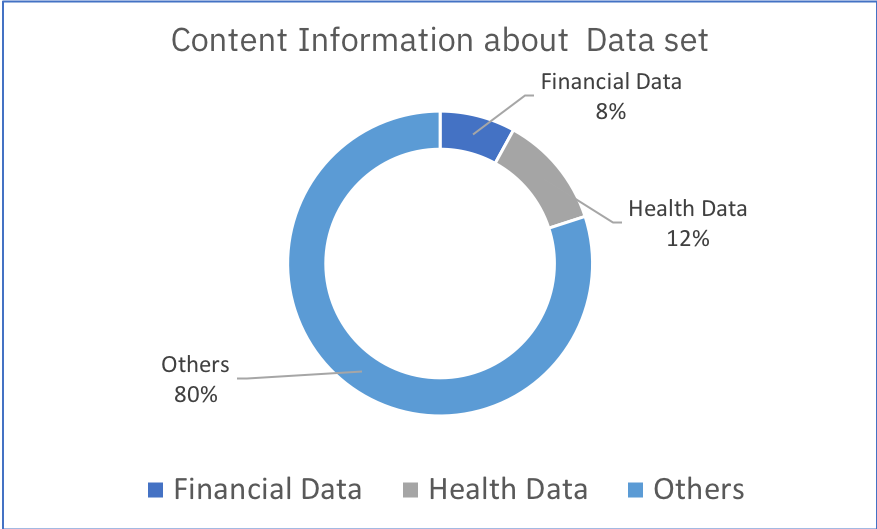}
    \caption{Content Information}
    \label{fig:structured}
  \end{subfigure}
   \begin{subfigure}[b]{0.29\textwidth}
    \includegraphics[scale=0.25]{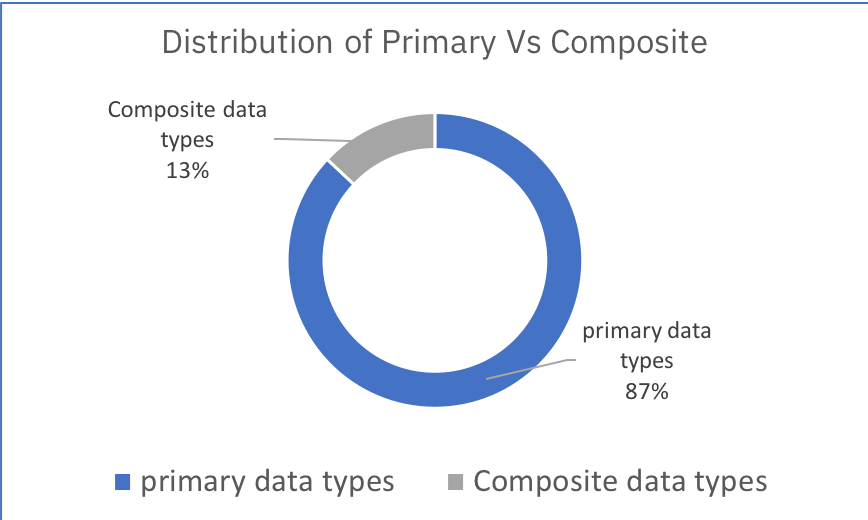}
    \caption{Primary Vs composite Data Types}
    \label{fig:primary}
  \end{subfigure}
  \\
     \begin{subfigure}[b]{0.29\textwidth}
    \includegraphics[scale=0.25]{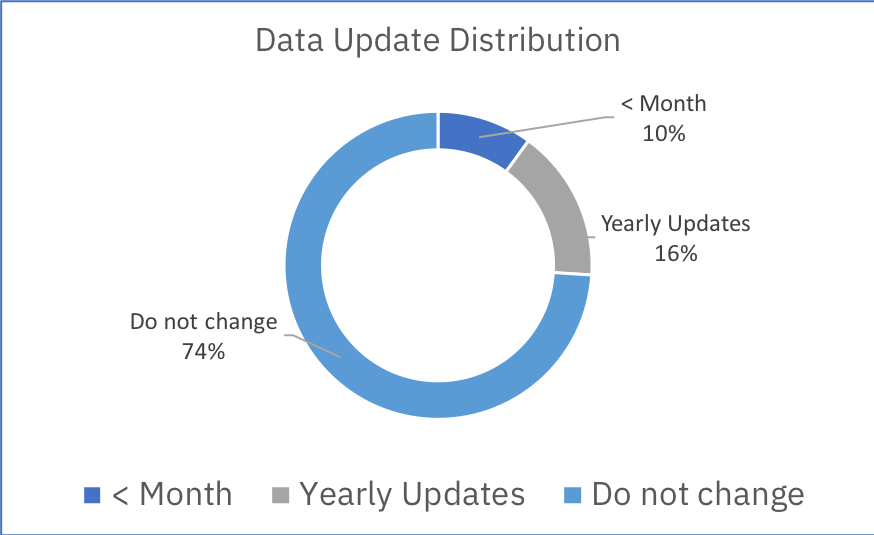}
    \caption{Frequency of update}
    \label{fig:frequency}
  \end{subfigure}
     \begin{subfigure}[b]{0.29\textwidth}
    \includegraphics[scale=0.25]{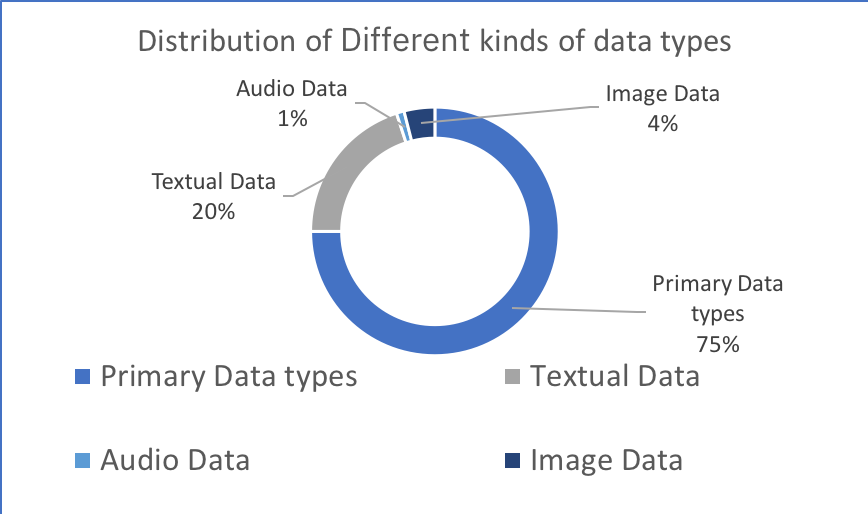}
    \caption{Data Types}
    \label{fig:datatype}
  \end{subfigure}
     \begin{subfigure}[b]{0.29\textwidth}
    \includegraphics[scale=0.25]{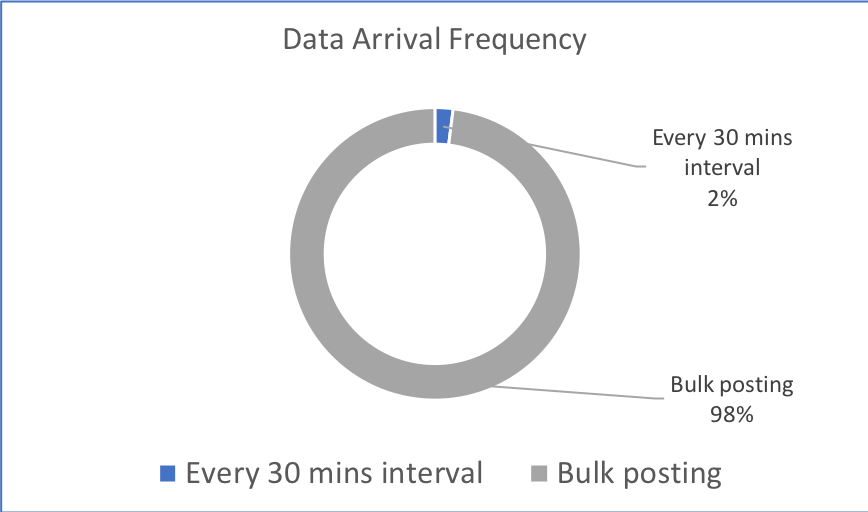}
    \caption{Data Arrival Frequency}
    \label{fig:arrival}
  \end{subfigure}
  \caption{Distribution of different facets/sub-facets studied using ~10500 different data sets from Kaggle \label{fig:distribution}}
  \vspace{-0.2in}
\end{figure*}


We begin with our hypothesis that the facets and sub-facets which comprise the metadata, form a basis for converting the qualitative description of the data to objective measures. Each sub-facet can take a set of values and the main question is to provide a mechanism to understand the preferred values for the facets, and ranking among the values for each facet, such that the ranks can be converted to logical scores.  The approach we have taken is to assume a preferred ordering for the various values of each sub-facet and convert this to a numerical score. We study the metadata of data sets in two well-known repositories to understand how the various values for each of the facets are distributed across the data sets in the two repositories. To begin with, this would form a prior for the values.


We first studied the distribution of metadata values on the data sets available on Kaggle~\cite{kaggle}.
Kaggle exposes data sets, metadata and kernels developed on those data sets, making it a good candidate to study the possible values for each sub-facet and their distribution across the data sets in the repository, for the classes of kernels (A kernel may be viewed as a specific application developed on the data set.) The distribution study influences both the value (or range of values) that data sets can take  and  how values should be ranked for assigning quantitative measurable value for each facet. At the time of these experiments, there were about  10550 data sets; each data set exposes meta data information such as tags, kernels and file types. Further each data set has 2-3 kernels leading to ~30000 kernels (ML algorithms) developed using the data sets. The meta data for each data set, comprising among other things the tags, title, identifier, size, schema and format, and the list of kernels for each data set were downloaded from Kaggle using their API \cite{kaggleapi} and the distribution was studied for each of the facets, across the data sets.

Figure \ref{fig:distribution} shows the different facets and the distribution of the values according to the number of data sets that has the specific values. Given the metadata information from Kaggle we were able to study twelve facets/sub facets from the data and data meta data directly. For large number of facets, the distribution indicates the nature of data sets and can be mapped to the preferred choice for users. From the figure it can be seen that the distribution for the values of the various attributes, that include data types, PII information, format, protected attributes availability, data size, schema availability, data level, structured vs unstructured, primary vs compsite, frequency of update can be used to derive a preferential order.   For some facets/sub-facets, the value is intrinsic to data and can be directly obtained from the data set owner. For example, facets related to data quality and statistical properties are derivable by nature of definitions described and assessing the data against them. For some facets, the observations on kaggle data set are uni-valued. For example all data sets in the kaggle are OpenSource, downloadable and easily processed through standard querying tools. Table \ref{table:distribution} shows the ranking of the values as assessed through Figure \ref{fig:distribution}. Facets such as Data Size and Data Content show distributions which differ from the preferred choice for these attributes. For instance, according to the frequency, data sets of size $<500 KB$ are more in number, but typically ML algorithms prefer larger data sets for learning more patterns. Similar study needs to be performed for domain verticals and context such as enterprises etc., before deriving the applicable values and the ranking for that domain/context.

\begin{table}[]
\scriptsize
    \centering
    \begin{tabular}{|c|l|l|}
    \hline
        Data Facet  & \parbox{0.75in}{Probable Values} & \parbox{1.25in}{Ranking (in decreasing order of choice)}  \\
        \hline
         PII information & Y/N  & N,Y\\
         Format & \parbox{0.75in}{CSV, TSV, SQLite, PDF, JSON, BigQuery} &\parbox{1.25in}{CSV,JSON, BigQuery, SQLite, PDF, TSV}\\
         Protected Attributes & Y/N & N,Y\\
         Data Size & \parbox{1in} {$1-500 KB, 500 KB - 500 MB, 500 MB- 1GB, 1 GB- 10 GB, >10 GB$} & \\
         Data schema &  Y/N & Y,N \\
         Data Level & Aggregate,Individual & Individual,Aggregate\\
         Data Layout & Structured,Unstructured & Structured,Unstructured\\
         Data Aggregation type & Primary, Composite &Primary,Composite\\
         Frequency of Update  & \parbox{0.75in}{Updated once, $<month$, yearly} & \parbox{1in}{Once, Yearly, Monthly}\\
         Data Type & \parbox{0.75in}{Primary data types, Audio,Image, Textual} & \parbox{1.25in}{Primary data types, Textual,Image,Textual}\\
         Data Arrival frequency & Less Frequent, Frequent & Less Frequent,Frequent\\
         \hline
    \end{tabular}
    \caption{Relative ordering of  values for different facets, based on the observed distributions}
    \vspace{-0.3in}
    \label{table:distribution}
    
\end{table}

\begin{figure*}[]
\centering
  \begin{subfigure}[b]{0.29\textwidth}
    \includegraphics[scale=0.25]{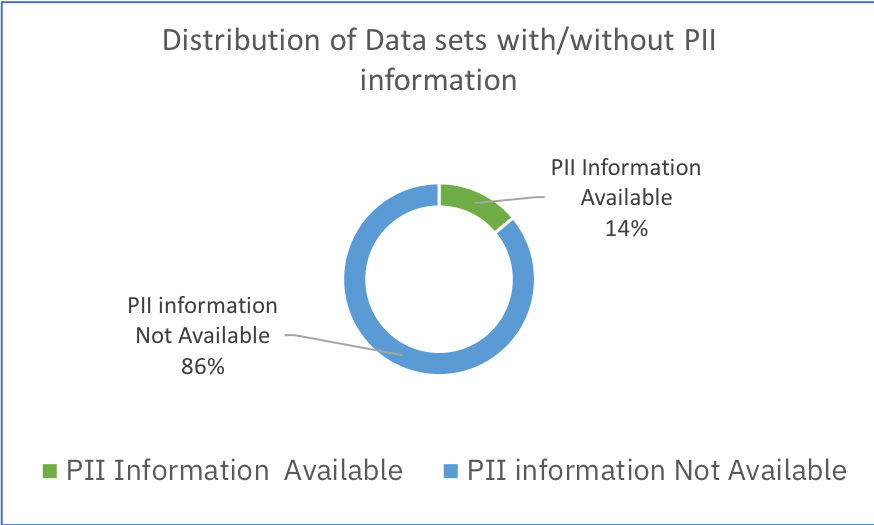}
    \caption{PII Distribution}
    \label{fig:ucipii}
  \end{subfigure}
  \begin{subfigure}[b]{0.29\textwidth}
    \includegraphics[scale=0.25]{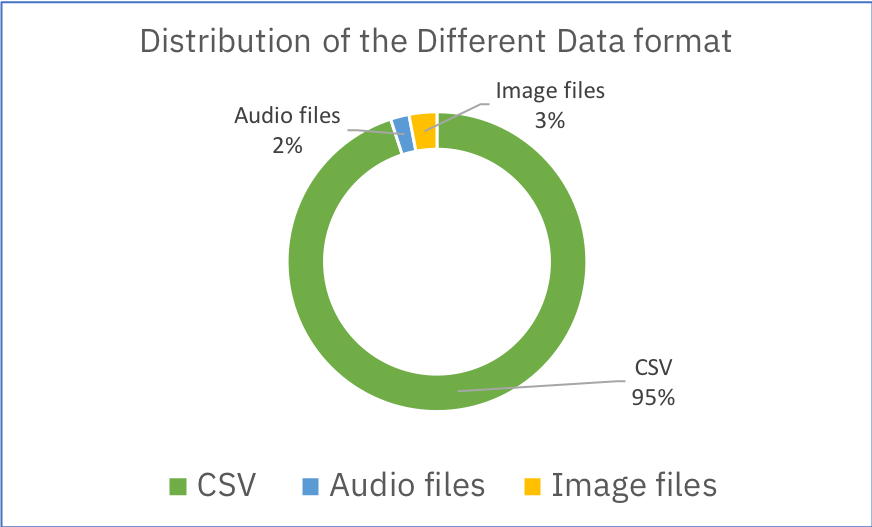}
    \caption{Format Distribution}
    \label{fig:uciformat}
  \end{subfigure}
\begin{subfigure}[b]{0.29\textwidth}
    \includegraphics[scale=0.25]{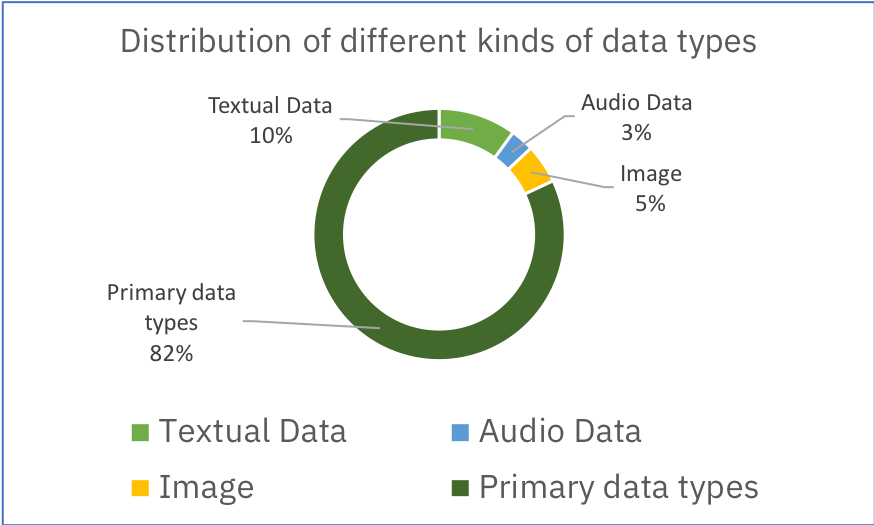}
    \caption{Data Type Distribution}
    \label{fig:ucibias}
  \end{subfigure}
  \\
  \begin{subfigure}[b]{0.29\textwidth}
    \includegraphics[scale=0.25]{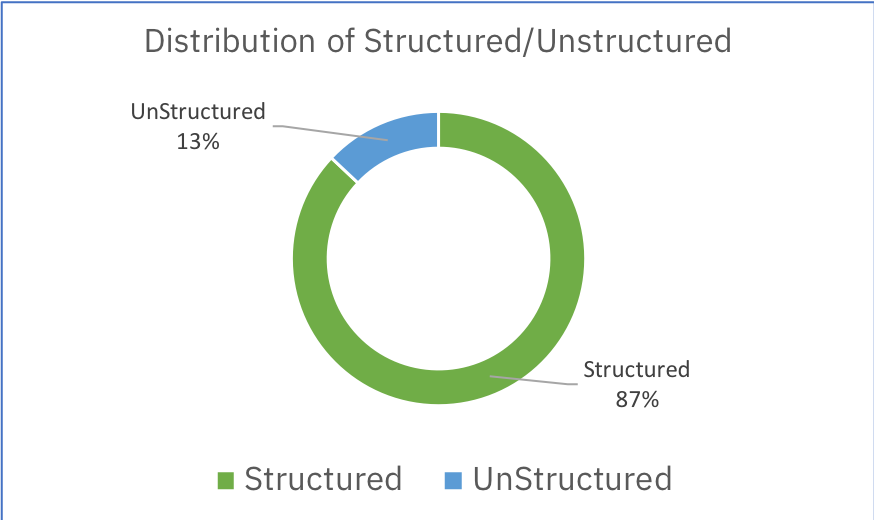}
    \caption{Structured Vs Unstructured}
    \label{fig:ucidatasize}
  \end{subfigure}
   \begin{subfigure}[b]{0.29\textwidth}
    \includegraphics[scale=0.25]{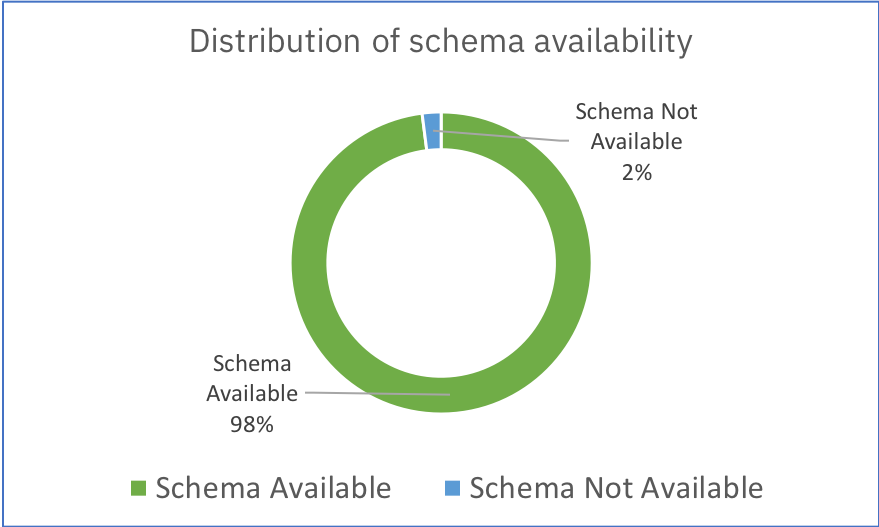}
    \caption{Schema Availability}
    \label{fig:ucischema}
  \end{subfigure}
\begin{subfigure}[b]{0.29\textwidth}
    \includegraphics[scale=0.25]{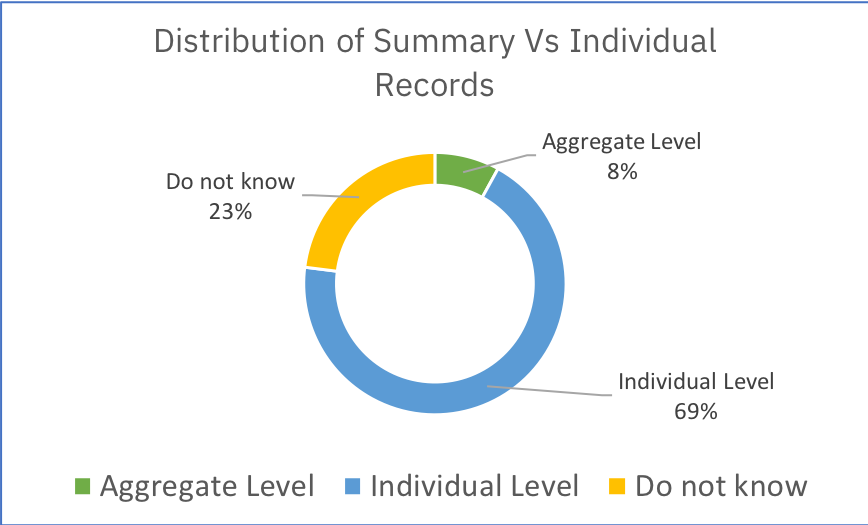}
    \caption{Data Level}
    \label{fig:ucisummary}
  \end{subfigure}
  \caption{Distribution of different facets/sub-facets studied using ~148 data sets from UCI repository for the task `classification' \label{fig:ucidistribution}}
  \vspace{-0.2in}
\end{figure*}

 The second repository in our study comprised the data sets available on the UCI site~\cite{ucidataset}. Here, along with the metadata on data types, schema, format, summary, identifier fields and data level, information on the different tasks (classification, clustering, regression) was also available. Figure \ref{fig:ucidistribution} presents a limited set of facets studied for the task `classification'.  From these studies, we derive a set of prior for values and ranks  based on the usage pattern.
 


\section{From data properties to data value}
\label{system}

In this section we present an overview of our system and then describe in detail the process of evaluation of a data set, by listing the questions, assigning scores for possible responses and evaluating these scores. 
\subsection{High-level view of the system}
\label{systemOverview1}

\begin{figure}[!h]
    \centering
    \includegraphics[scale=0.25]{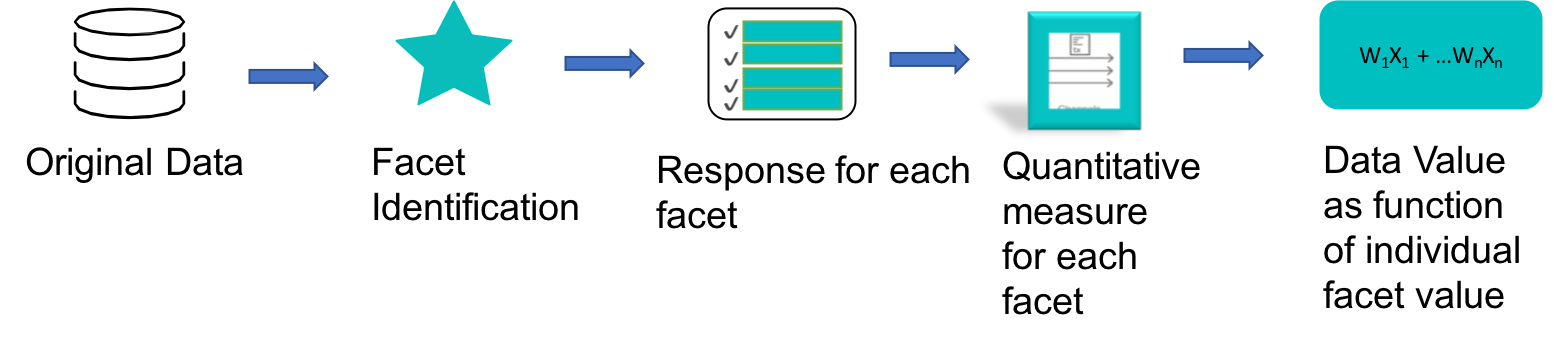}
    \caption{System showing computing data value from facets}
    \label{fig:systemoverview}
    \vspace{-0.2in}
\end{figure}

Figure \ref{fig:systemoverview} provides a schematic diagram of the various steps in assessing the data value categorically.

We first enumerate the various facets of the data, that describe
different aspects and may have an impact on the data value in some manner or the other. We have tried to be exhaustive in our listing and have also drawn upon existing literature that has looked at data characteristics in other contexts. 
For each of the facets we draw up a set of questions that can be  answered by the person who has access to the data set, whom we will refer to as the respondent.

The questions have been designed with the following in mind.
\begin{enumerate}[wide, labelwidth=!, labelindent=0pt]
\item Each question has an objective response - the response is either binary, or quantitative, or from a set of categorical values. 
\item  The respondent can independently answer the questions only with the knowledge of the data set and no information about the target applications. 
\item The questions are designed to address all the various minute details of different facets that could impact value. Some questions may appear to have an overlap and this is done to include completeness at the cost of repetition.
\item By enumerating a list of values for each sub-facet, the respondent can identify transformations to the data that might make it more valuable.
\item The questionnaire in this format can be extended on different fronts, from respondents in different roles, to make it more exhaustive and hence more rigourous.

\end{enumerate} 

Not all questions will be applicable to all data sets, in which case
the responses can be omitted. Each facet represents a certain characteristic of the data, and has associated a set of sub-facets. Further, the role of the respondent will influence the value assigned as response for each question. Broadly we envision two roles, the seller and the buyer, though there could be many more. The facets are also not mutually exclusive in some cases. Some of the questions could be mutually exclusive, in which case only one response is considered in the quantitative assessment.

\subsection{Identifying values for  Facets/Sub-Facets}

The response to each question is from a set of values which could be binary, categorical or a numerical value between $0$ and $1$. For each question that the respondent thinks is relevant to the data set, he or she selects the value applicable. Some questions may be mutually exclusive. 
The high-level process is shown in Figure \ref{fig:systemoverview}; the questions are defined, possible responses for each are defined and an ordinal ranking is defined for the possible values, for each question. This  ranking is best defined when the context is known. However, even in the absence of context, there is consensus on some of the characteristics, in terms of what is usually preferred across users in general. An example is, in general more recent data is preferred to less recent data, and data without any legal restrictions is preferred to data with legal restrictions. For binary-valued responses, the more preferred response is given a value of $1$ while the other response is given a value of $0$.  In the case of categorical values, a value may be assigned based on the ordering, as for instance, if the allowed responses are \emph{Large}, \emph{Medium} and \emph{Small}, and \emph{Large} is most preferred, then this may be assigned a value 1, while \emph{Medium} may be assigned a value of $0.5$ and \emph{Small} may be assigned a value of $0$. Similarly, in some cases the respondent may be asked to assign a value between 0 and 1, for the specific attribute, such as data accuracy. Further, in some questions requiring a quantitative response, if the respondent is unable to specify a numeric value, the questionnaire may include an additional set of options that are categorical, and the respondent may choose to respond to this question.

\subsection{Deriving the data value}

 Let $V_{i}$ represent the value associated with the selected response for the $i^{th}$ question. The following formula is then used to derive the data value.
 
\begin{equation}
DataValue \leq \sum_{i=1}^{n} \alpha_{i} V_{i}, \left\{ \begin{array}{rcl}
0 \leq V_i \leq 1\\
\sum_1^n \alpha_{n} =1
\end{array}\right.  \\
\label{eqn:dv}
\end{equation}

where $V_i$ is the value selected as response for the $i^{th}$ question, which could be any value between and including $0$ and $1$, and $\alpha_{i}$ represents the relative weights of the question for any data set. For example, the availability of a schema for a data set will add more value to the data set compared to whether the data set volume exceeds a certain limit. Equation \ref{eqn:dv} evaluate the data value on a  score that could range from $0$ to a maximum of $n$, where $n$ is the number of questions. Further, from the end-user's point of view, equation \ref{eqn:dv}  may be viewed as a multi-attribute utility value of the data, where each characteristic is the attribute, and $V_i$ is the perceived utility of that sub-facet. 
The quantification now provides a basis for comparing data sets for applications, identifies the most relevant data set for the application and can be extended to enable searching for data set in large warehouses based on facets and sub-facets.




\subsection{Sub-facets in detail}
\label{properties}

Tables \ref{tab:allFacets1},\ref{tab:allFacets2},\ref{tab:allFacets3} and \ref{tab:allFacets4}  capture a top-level description of the sub-facets and a set of questions that relate to it.\footnote{The ordering of the facets in the tables has no significance - it has mainly been done to optimize on space usage.} We have tried to list the facets and sub-facets as exhaustively as possible. However we view this list as only a starting set, which will be extended in due course by other users, as more facets are identified as relevant for different applications. The terms themselves may be defined differently across different usage scenarios; we use the definition that we consider most relevant in the context of data value assessment. 


 Each question has a set of possible values, from which the respondent selects the value applicable for that data set. Each value is assigned a score, based on an assumed preferences. The combined scoring across all the facets as defined in equation~\ref{eqn:dv} then yields a quantification of data value under these set of assumptions, which may be used for comparing with other data sets under the same set of assumptions, or for further fine-tuning when more context is available. Note that in the following, where the question has a binary response, a 'Y' or a 'N', the score is 1 for a 'Y' and 0 for a 'N' unless indicated otherwise. Further, a 'Dont know' option is possible for many questions, and this has not been explicitly shown here; when selected, it would have a score of 0.

\begin{table}[]
    \centering
    \begin{tabular}{|p{85mm}|}
    \hline
    \multicolumn{1}{|c|}{Data Layout}  \\
    \hline
      {Structured data is any data that has well defined boundaries, in the form as it is, and that can be used to identify data points, fields and instances. 

Unstructured data is data which, in the form given, does not have well defined boundaries to identify  data points  or instances. Examples include  binary data, blobs of text, video, audio and image files.

Semi-structured data is any data which in the given form has a structure, but where the content within the boundaries could contain unstructured data. Examples are XML and HTML documents.}\\
\multicolumn{1}{|c|} {\textasciitilde \textasciitilde \textasciitilde}\\
a) What is the data structure?  \\ \mbox{\ooalign{Structured}}/\mbox{\ooalign{Semi-structured}}/\mbox{\ooalign{Unstructured}}\\ 
The respective scores for the above values would be 1.0, 0.5, and 0.25\\
\hline
 \multicolumn{1}{|c|}{Data age}\\
 \hline
 In general we would state that more recent data is better than less recent data. Further, some data gets outdated more rapidly than others, and hence in general, if the data is relevant for a longer period, it is more useful.\\
\multicolumn{1}{|c|} {\textasciitilde \textasciitilde \textasciitilde}\\
a) How current is the data? 
\mbox{\ooalign{Latest}}/\mbox{\ooalign{Recent}}/\mbox{\ooalign{Old}}/\mbox{\ooalign{Dont know}}\\ 
Scores: 1, 0.75, 0.25 and 0 respectively\\
b) Is there a known later instance of the data? \mbox{\ooalign{y\cr\hidewidth$\square$\hidewidth\cr}}/\mbox{\ooalign{n\cr\hidewidth$\square$\hidewidth\cr}}\\
Score: 0 and 1 respectively\\
c) How frequently does the data get outdated/updated?\\
\mbox{\ooalign{Daily}}/\mbox{\ooalign{Weekly}}/\mbox{\ooalign{Monthly}}/\mbox{\ooalign{Yearly}}/\mbox{\ooalign{Not applicable}}/\mbox{\ooalign{Dont know}}\\
Score: 0.25, 0.5,0.75, 1.0, 0 and 0 respectively\\
\hline
\multicolumn{1}{|c|}{Data volume}\\
 \hline
 This is the size of the data set, in KB,MB or other relevant unit, and would have a bearing on the storage and processing costs. Modulo a certain upper bound, more data is generally better than less data.\\
\multicolumn{1}{|c|} {\textasciitilde \textasciitilde \textasciitilde}\\
a) What is the data size?\\
\mbox{\ooalign{size$<$500M;}}\mbox{\ooalign{500MB$\leq$ size$<$ 10G;}}\\
\mbox{\ooalign{10G~$<$size~$\leq$~100G;}}\mbox{\ooalign{size$>$ 100G}}\\
Score: 0.5, 0.75, 1.0 and 0.5 respectively.\\
\hline
\multicolumn{1}{|c|}{Composition of the data }\\
 \hline
 This captures information on level of homogeneity in the data.\\
 \multicolumn{1}{|c|} {\textasciitilde \textasciitilde \textasciitilde}\\
a) Are the data point instances primary data type? \mbox{\ooalign{y\cr\hidewidth$\square$\hidewidth\cr}}/\mbox{\ooalign{n\cr\hidewidth$\square$\hidewidth\cr}}\\
b) Are all instances  similar?~\mbox{\ooalign{y\cr\hidewidth$\square$\hidewidth\cr}}/\mbox{\ooalign{n\cr\hidewidth$\square$\hidewidth\cr}}\\
\hline
\multicolumn{1}{|c|}{Data Granularity}\\
\hline
Aggregate information is generally less useful than detailed or instance-level information. \\
\multicolumn{1}{|c|} {\textasciitilde \textasciitilde \textasciitilde}\\
a) Is it aggregate or summary information?
\mbox{\ooalign{y\cr\hidewidth$\square$\hidewidth\cr}}/\mbox{\ooalign{n\cr\hidewidth$\square$\hidewidth\cr}}\\
Score: 0, 1\\
\hline

 \multicolumn{1}{|c|}{Data Usage}\\
 \hline
Data that is easier to use is preferred to data that is more complex to use. These aspects are also covered in sub-facets of format and type. However this facet is being retained because in some instances, the respondent may have a specific view on this aspect of the data.\\
\multicolumn{1}{|c|} {\textasciitilde \textasciitilde \textasciitilde}\\
a) How easy is it to utilise the data? 
~\mbox{\ooalign{Simple}}/\mbox{\ooalign{Moderate}}/\mbox{\ooalign{Difficult}}/\mbox{\ooalign{Complex}}\\
Score: 1, 0.6, 0.3, 0 respectively\\
\hline

 \end{tabular}
    
    \caption{Sub-facets: Set 1\label{tab:allFacets1}}
\end{table}

\begin{table}[]
    \centering
    \begin{tabular}{|p{85mm}|}
    \hline
    \multicolumn{1}{|c|}{Data Format}  \\
    \hline
This covers a set of sub-facets as follows; the format of the file is the format in which the information in the file is stored - common examples being csv, pdf or gif. Some formats in general have been more widely used with greater support for interoperability and processing.  In the case of structured data, the data set has a
schema if, apart from the data, there is also metadata that tells us something about the
relationship of the different data items. It is relational data if it can be presented
in tabular form or moved to a relational database. In the case of standards, it is possible to check the data set against a known set of standards and understand compliance with any of them. Compliance to standards may imply availability of generic tools and query languages to operate in the data. A proprietary format is a format that is owned by someone or some entity; however, as long as it is open, it will not be a restriction. \\
\multicolumn{1}{|c|} {\textasciitilde \textasciitilde \textasciitilde}\\
    
a) What is the format of the data set file?\\
\mbox{\ooalign{csv}}/\mbox{\ooalign{pdf}}/\mbox{\ooalign{tsv}}/\mbox{\ooalign{gif,jpg}}/\mbox{\ooalign{xml}}/\mbox{\ooalign{json}}/\mbox{\ooalign{other}}\\
Score: 1 for xml, json, csv, tsv; 0 for pdf, gif, jpg, other\\
b) Does it have a schema?~\mbox{\ooalign{y\cr\hidewidth$\square$\hidewidth\cr}}/\mbox{\ooalign{n\cr\hidewidth$\square$\hidewidth\cr}}\\
Score: 1 and 0 respectively, for this response, and for other binary-valued responses.\\
c) Is it an export or query result of relational data? ~\mbox{\ooalign{y\cr\hidewidth$\square$\hidewidth\cr}}/\mbox{\ooalign{n\cr\hidewidth$\square$\hidewidth\cr}}\\
d) Does it adhere to a standard?~\mbox{\ooalign{y\cr\hidewidth$\square$\hidewidth\cr}}/\mbox{\ooalign{n\cr\hidewidth$\square$\hidewidth\cr}}\\
e) If in proprietary format, is it open? \mbox{\ooalign{y\cr\hidewidth$\square$\hidewidth\cr}}/\mbox{\ooalign{n\cr\hidewidth$\square$\hidewidth\cr}}\\
f) Is it in normalized form?~\mbox{\ooalign{y\cr\hidewidth$\square$\hidewidth\cr}}/\mbox{\ooalign{n\cr\hidewidth$\square$\hidewidth\cr}}\\
\hline

\multicolumn{1}{|c|}{Data Sensitivity}\\
\hline
This aspect could influence the data cost significantly in terms of the added risk of exposure, and need to protect the data from breaches. However it would also increase the value of the data in terms of the improved analysis that is possible. For our purpose here, we assume that if the same task can be performed with data having sensitive information and data that does not have sensitive information, then  the latter would be preferred. While the legal aspects are covered as a separate facet, there could be additional moral and ethical issues in the dissemination of such data.\\
\multicolumn{1}{|c|} {\textasciitilde \textasciitilde \textasciitilde}\\
a) Is it free of confidential information? ~\mbox{\ooalign{y\cr\hidewidth$\square$\hidewidth\cr}}/\mbox{\ooalign{n\cr\hidewidth$\square$\hidewidth\cr}}\\
b) Is it free of personal identifiable information? ~\mbox{\ooalign{y\cr\hidewidth$\square$\hidewidth\cr}}/\mbox{\ooalign{n\cr\hidewidth$\square$\hidewidth\cr}}\\
c) Is it free of information to be retained for mandatory purposes? ~\mbox{\ooalign{y\cr\hidewidth$\square$\hidewidth\cr}}/\mbox{\ooalign{n\cr\hidewidth$\square$\hidewidth\cr}} \\
d) Is it free of revenue or financial data? ~\mbox{\ooalign{y\cr\hidewidth$\square$\hidewidth\cr}}/\mbox{\ooalign{n\cr\hidewidth$\square$\hidewidth\cr}}\\
e) Does it have Medical data/ health data? ~\mbox{\ooalign{y\cr\hidewidth$\square$\hidewidth\cr}}/\mbox{\ooalign{n\cr\hidewidth$\square$\hidewidth\cr}}\\
Score: 0 and 1 respectively\\
f) Does it have protective variables? ~\mbox{\ooalign{y\cr\hidewidth$\square$\hidewidth\cr}}/\mbox{\ooalign{n\cr\hidewidth$\square$\hidewidth\cr}}\\
Score: 0 and 1 respectively\\
\hline

\multicolumn{1}{|c|} {Data Velocity}\\
\hline
This is the rate at which the data arrives. It influences the design of the data store systems and the plans for
scalability. Specifically for streaming it requires computational resources.\\
\multicolumn{1}{|c|} {\textasciitilde \textasciitilde \textasciitilde}\\
a) How rapidly can it be said to be generated? \\
\mbox{\ooalign{Very fast}}/\mbox{\ooalign{Fast}}/\mbox{\ooalign{Medium}}/\mbox{\ooalign{Not significant}}\\
Score: 0.5, 0.75, 1.0 and 1.0 respectively\\
b) Is it streaming data?
\mbox{\ooalign{y\cr\hidewidth$\square$\hidewidth\cr}}/\mbox{\ooalign{n\cr\hidewidth$\square$\hidewidth\cr}} \\
\hline
 \multicolumn{1}{|c|}{Data processing}\\
 \hline
Presence of tools to read and analyse the data makes it more useful than otherwise.\\
\multicolumn{1}{|c|} {\textasciitilde \textasciitilde \textasciitilde}\\
a) Are there tools or programs to cleanse the data? \mbox{\ooalign{y\cr\hidewidth$\square$\hidewidth\cr}}/\mbox{\ooalign{n\cr\hidewidth$\square$\hidewidth\cr}}\\
b) Are there tools or programs to process the data in the current format? \mbox{\ooalign{y\cr\hidewidth$\square$\hidewidth\cr}}/\mbox{\ooalign{n\cr\hidewidth$\square$\hidewidth\cr}}\\
\hline
 \end{tabular}
    \caption{Sub-facets: Set 2 \label{tab:allFacets2}}
\end{table}

\begin{table}[]
    \centering
    \begin{tabular}{|p{85mm}|}
    \hline
    \multicolumn{1}{|c|}{Statistical properties}  \\
    \hline
    
These properties are important intermediate steps in the path to insights from data. Hence, though they may be hard to define, we are listing them here.\\
\multicolumn{1}{|c|} {\textasciitilde \textasciitilde \textasciitilde}\\

a) Is it suitable for classification models? ~\mbox{\ooalign{y\cr\hidewidth$\square$\hidewidth\cr}}/\mbox{\ooalign{n\cr\hidewidth$\square$\hidewidth\cr}}\\
b) Is it suitable for linear regression models? ~\mbox{\ooalign{y\cr\hidewidth$\square$\hidewidth\cr}}/\mbox{\ooalign{n\cr\hidewidth$\square$\hidewidth\cr}}\\
c) Is it suitable for clustering models?~\mbox{\ooalign{y\cr\hidewidth$\square$\hidewidth\cr}}/\mbox{\ooalign{n\cr\hidewidth$\square$\hidewidth\cr}}\\
d) Has it been used in ML algorithms already?~\mbox{\ooalign{y\cr\hidewidth$\square$\hidewidth\cr}}/\mbox{\ooalign{n\cr\hidewidth$\square$\hidewidth\cr}}\\
e) Was any sampling applied on the data to get this sample? ~\mbox{\ooalign{y\cr\hidewidth$\square$\hidewidth\cr}}/\mbox{\ooalign{n\cr\hidewidth$\square$\hidewidth\cr}}\\
f) Is it time-series data?~\mbox{\ooalign{y\cr\hidewidth$\square$\hidewidth\cr}}/\mbox{\ooalign{n\cr\hidewidth$\square$\hidewidth\cr}}\\
g) Is it suitable for bivariate analysis? \mbox{\ooalign{y\cr\hidewidth$\square$\hidewidth\cr}}/\mbox{\ooalign{n\cr\hidewidth$\square$\hidewidth\cr}}\\
h) Is it suitable for multivariate analysis? \mbox{\ooalign{y\cr\hidewidth$\square$\hidewidth\cr}}/\mbox{\ooalign{n\cr\hidewidth$\square$\hidewidth\cr}}\\

\hline

\multicolumn{1}{|c|} {Frequency of Use}\\
\hline
The current frequency of use is a rough indicator of future use or disuse of the data.\\
\multicolumn{1}{|c|} {\textasciitilde \textasciitilde \textasciitilde}\\
When was the data last used? \\
~\mbox{\ooalign{This month}}/\mbox{\ooalign{This year}}/ \mbox{\ooalign{In last 5 years}}/\mbox{\ooalign{More than 5 years ago}}\\
Score: 1, 0.75, 0.5 and 0 respectively\\
Is there a 'known' future use? \mbox{\ooalign{y\cr\hidewidth$\square$\hidewidth\cr}}/\mbox{\ooalign{n\cr\hidewidth$\square$\hidewidth\cr}}\\
\hline

\multicolumn{1}{|c|} {Data Quality}\\
\hline
This is one aspect of data that has been quite well-studied in quantitative terms, for its impact on the data value. 
We pose the following questions categorically, to be answered as best as possible as it purports to the current definition of the data set. A list of values may be provided for selection, in case the respondent is unable to provide a specific number as response.\\
\multicolumn{1}{|c|} {\textasciitilde \textasciitilde \textasciitilde}\\
a) Are all the fields complete?~\mbox{\ooalign{y\cr\hidewidth$\square$\hidewidth\cr}}/\mbox{\ooalign{n\cr\hidewidth$\square$\hidewidth\cr}}?\\
b) Is it error-free?~ \mbox{\ooalign{y\cr\hidewidth$\square$\hidewidth\cr}}/\mbox{\ooalign{n\cr\hidewidth$\square$\hidewidth\cr}}\\
c) Are there known missing instances? \mbox{\ooalign{y\cr\hidewidth$\square$\hidewidth\cr}}/\mbox{\ooalign{n\cr\hidewidth$\square$\hidewidth\cr}}\\
Score: 0 and 1 respectively\\
d) Does it fill the missing values in an existing data set? ~\mbox{\ooalign{y\cr\hidewidth$\square$\hidewidth\cr}}/\mbox{\ooalign{n\cr\hidewidth$\square$\hidewidth\cr}}\\
e) Is it complete, with respect to the purpose it defines? ~\mbox{\ooalign{y\cr\hidewidth$\square$\hidewidth\cr}}/\mbox{\ooalign{n\cr\hidewidth$\square$\hidewidth\cr}}\\
f) Is it known to have duplicates?~\mbox{\ooalign{y\cr\hidewidth$\square$\hidewidth\cr}}/\mbox{\ooalign{n\cr\hidewidth$\square$\hidewidth\cr}}\\
Score: 0 and 1 respectively\\
g) Does it complement or supplement an existing data set?~\mbox{\ooalign{y\cr\hidewidth$\square$\hidewidth\cr}}/\mbox{\ooalign{n\cr\hidewidth$\square$\hidewidth\cr}}\\
h) Is the data accurate?~\mbox{\ooalign{y\cr\hidewidth$\square$\hidewidth\cr}}/\mbox{\ooalign{n\cr\hidewidth$\square$\hidewidth\cr}}\\
i) What is the precision? \\
Enter a number between 0 and 1 or select one of: ~\mbox{\ooalign{High}}/\mbox{\ooalign{Medium}}/\mbox{\ooalign{Low}}\\
j) What is the recall? \\
Enter a number between 0 and 1 or select one of: ~\mbox{\ooalign{High}}/\mbox{\ooalign{Medium}}/\mbox{\ooalign{Low}}\\
Score: Numeric value or 1,0.5 and 0 respectively\\
k) Is the data consistent within the data set? ~\mbox{\ooalign{y\cr\hidewidth$\square$\hidewidth\cr}}/\mbox{\ooalign{n\cr\hidewidth$\square$\hidewidth\cr}}\\
Score: Numeric value or 1,0.5 and 0 respectively\\
l) Does the data have noise?~\mbox{\ooalign{y\cr\hidewidth$\square$\hidewidth\cr}}/\mbox{\ooalign{n\cr\hidewidth$\square$\hidewidth\cr}}\\
Score: 0 and 1 respectively\\
\hline

\multicolumn{1}{|c|}{Transformation on the data}\\
\hline
Transformations on the data indicate that some level of processing has already been done on the data in order to make it more consumable. Some transformations such as data anonymization may reduce the potential for some kinds of personalized analysis. However, in the absence of context, we posit that anonymized data, which can be shared more readily, is preferred to non-anonymized data.\\
 \multicolumn{1}{|c|} {\textasciitilde \textasciitilde \textasciitilde}\\
Is it known to have had data transformations?\mbox{\ooalign{y\cr\hidewidth$\square$\hidewidth\cr}}/\mbox{\ooalign{n\cr\hidewidth$\square$\hidewidth\cr}} \\
Is it encrypted data?\mbox{\ooalign{y\cr\hidewidth$\square$\hidewidth\cr}}/\mbox{\ooalign{n\cr\hidewidth$\square$\hidewidth\cr}}\\
Is it anonymized data? \mbox{\ooalign{y\cr\hidewidth$\square$\hidewidth\cr}}/\mbox{\ooalign{n\cr\hidewidth$\square$\hidewidth\cr}}\\
\hline

 \end{tabular}
    
    \caption{Sub-facets: Set 3\label{tab:allFacets3}}
\end{table}
   
\begin{table}[]
    \centering
    \begin{tabular}{|p{85mm}|}
    \hline
    \multicolumn{1}{|c|}{Data sourcing}  \\
    \hline
    The source of the data, including the purpose of creation of the data, again bear on the cost of generation, ease of availability in general to others, whether there are alternate sources and how straightforward or not the data collection exercise is.  Single sources poses lesser overhead in terms of maintaining updates while multiple sources require synchronization from different sources to keep the data update if one portion gets updated.\\
    \multicolumn{1}{|c|} {\textasciitilde \textasciitilde \textasciitilde}\\
a) Can this data be easily accessed by all? ~\mbox{\ooalign{y\cr\hidewidth$\square$\hidewidth\cr}}/\mbox{\ooalign{n\cr\hidewidth$\square$\hidewidth\cr}} \\
Score: 0 and 1 respectively\\
b) How was the data obtained?\\
\mbox{\ooalign{Survey}}/\mbox{\ooalign{Customer feedback}}/\mbox{\ooalign{Transactional}}/\mbox{\ooalign{Web crawler}}//
\mbox{\ooalign{Licensing}}/\mbox{\ooalign{Outright purchase}}/\mbox{\ooalign{Others}}\\
Score: 1,1,0.5,0,0.5,0.75,0 respectively\\
c) Is the data aggregated from many sources or from single source?
~\mbox{\ooalign{Multiple sources}}/\mbox{\ooalign{Single source}}\\
d) Is this enterprise-generated?~\mbox{\ooalign{y\cr\hidewidth$\square$\hidewidth\cr}}/\mbox{\ooalign{n\cr\hidewidth$\square$\hidewidth\cr}}\\
e) Is this publicly available?~\mbox{\ooalign{y\cr\hidewidth$\square$\hidewidth\cr}}/\mbox{\ooalign{n\cr\hidewidth$\square$\hidewidth\cr}}
Score: 0 and 1 respectively.\\
f) Is this data machine generated?~
\mbox{\ooalign{y\cr\hidewidth$\square$\hidewidth\cr}}/\mbox{\ooalign{n\cr\hidewidth$\square$\hidewidth\cr}}\\
Score: 0 and 1 respectively.\\
g) Are there known alternates for this data set?
~\mbox{\ooalign{y\cr\hidewidth$\square$\hidewidth\cr}}/\mbox{\ooalign{n\cr\hidewidth$\square$\hidewidth\cr}}\\
Score: 0 and 1 respectively.\\
\hline
\multicolumn{1}{|c|} {Enterprise aspects}\\
\hline
The aspects of data here determine how it is perceived in the enterprise, in a general context.\\
    \multicolumn{1}{|c|} {\textasciitilde \textasciitilde \textasciitilde}\\
a) Is the data already making money? ~\mbox{\ooalign{y\cr\hidewidth$\square$\hidewidth\cr}}/\mbox{\ooalign{n\cr\hidewidth$\square$\hidewidth\cr}} \\
b) Will it improve the efficiency of an existing application or business process? ~\mbox{\ooalign{y\cr\hidewidth$\square$\hidewidth\cr}}/\mbox{\ooalign{n\cr\hidewidth$\square$\hidewidth\cr}} \\
c) Does it introduce a new channel to reach to customers?~\mbox{\ooalign{y\cr\hidewidth$\square$\hidewidth\cr}}/\mbox{\ooalign{n\cr\hidewidth$\square$\hidewidth\cr}}\\
d) Does it complement an existing application? ~\mbox{\ooalign{y\cr\hidewidth$\square$\hidewidth\cr}}/\mbox{\ooalign{n\cr\hidewidth$\square$\hidewidth\cr}}\\
e) Does it increase customer reach?~\mbox{\ooalign{y\cr\hidewidth$\square$\hidewidth\cr}}/\mbox{\ooalign{n\cr\hidewidth$\square$\hidewidth\cr}}\\
f) Which parts of the business process does it contribute to?\\
\mbox{\ooalign{Sales}}/\mbox{\ooalign{Marketing}}/\mbox{\ooalign{HR}}/\mbox{\ooalign{Operations}}/\mbox{\ooalign{Finance}}//
/\mbox{\ooalign{Accounting}}/\mbox{\ooalign{Payroll}}/\mbox{\ooalign{Others}}\\
Score: 0 for Others and 1 for everything else\\
g) At which hierarchy in the organization is the data used?
~\mbox{\ooalign{Executive}}/\mbox{\ooalign{Middle management}}/\mbox{\ooalign{Others}}/\mbox{\ooalign{Multiple}}\\
Score: 1,0.75, 0.5 and 1 respectively\\
\hline

\multicolumn{1}{|c|}{Legal and Access Aspects}\\
\hline
In a manner similar to the use of sensitive data, when a data set is unconditionally available for an application, it is preferred to a data set that may have legal controls or other restrictions operating on it. Ease of access in this context implies that there are not too many procedures or processes that make it difficult to start using the data.\\
 \multicolumn{1}{|c|} {\textasciitilde \textasciitilde \textasciitilde}\\
a) Is this data free of any legal restrictions in usage?~\mbox{\ooalign{y\cr\hidewidth$\square$\hidewidth\cr}}/\mbox{\ooalign{n\cr\hidewidth$\square$\hidewidth\cr}} \\
b) Was this data acquired as part of some contract? ~\mbox{\ooalign{y\cr\hidewidth$\square$\hidewidth\cr}}/\mbox{\ooalign{n\cr\hidewidth$\square$\hidewidth\cr}}\\
Score: 0 and 1 respectively\\
c) Are there any contractual obligations on the data? ~\mbox{\ooalign{y\cr\hidewidth$\square$\hidewidth\cr}}/\mbox{\ooalign{n\cr\hidewidth$\square$\hidewidth\cr}}\\
Score: 0 and 1 respectively\\
d) If pertaining to 'information about people', was there consent to use?~\mbox{\ooalign{y\cr\hidewidth$\square$\hidewidth\cr}}/\mbox{\ooalign{n\cr\hidewidth$\square$\hidewidth\cr}} \\
e) Is it governed by some license?~\mbox{\ooalign{y\cr\hidewidth$\square$\hidewidth\cr}}/\mbox{\ooalign{n\cr\hidewidth$\square$\hidewidth\cr}} \\
Score: 0 and 1 respectively\\
f) Are there export restrictions?~\mbox{\ooalign{y\cr\hidewidth$\square$\hidewidth\cr}}/\mbox{\ooalign{n\cr\hidewidth$\square$\hidewidth\cr}}\\
Score: 0 and 1 respectively\\
g) Is the data easy to access?~\mbox{\ooalign{y\cr\hidewidth$\square$\hidewidth\cr}}/\mbox{\ooalign{n\cr\hidewidth$\square$\hidewidth\cr}}\\
\hline

 \end{tabular}
    
    \caption{Sub-facets: Set 4\label{tab:allFacets4}}
\end{table}


\subsection{Relative ordering of facet values and facets}
\label{survey}


We need a method of establishing the preferred choice among the various values of a sub-facet, in order to achieve a scoring for each sub-facet. Our view is that any possible approach, either through extensive user surveys or through studying the distribution of these values from data sets used in developing various applications or algorithms (a sample of this is what is discussed in Section~\ref{assessment}, will not generalize at large. It is obviously not the case that the usage patterns observed for one class of applications or even a few classes can be generalized to score data values across all applications, or that usage patterns preferred by a select set of users would reflect usage patterns in general. However we indicate this as a method of getting started with a relative quantitative assessment for two or more data sets. As more context is available, it is straightforward to override the ordering of values and the valuation becomes more exact.



The relative priorities  $\alpha$ in Equation \ref{eqn:dv} for each of the facets may again be obtained in multiple ways - through a survey or a domain knowledge expert, for instance. In the simplest case, all facets may be assumed to have equal weight. As more information is available, the weights may be adjusted to reflect the needs of the actual application. When the application is available against which the data sets are evaluated, then the relative priority can be defined for the application requirements. For example, applications might require ONLY data at individual level and therefore all data sets have aggregate level information irrespective of matches for other facets are deemed unfit for the application.
\label{survey}


 \section{Illustration of Deriving Data Value}
\label{assessment}

We illustrate our technique of computing data value for data sets in two well-known repositories. Our goal is to show  both the simplicity of the questionnaire that can drive the data practitioner to instantly provide answers without much complexity and the feasibility of assigning a quantity measure that assesses the value of data. 

Our initial example is in the general context, without a specific application.  Consider a scenario where a data practitioner wants to search for prisons data set without any application context. To illustrate this scenario, we have used two data sets India Prisons Data Set \cite{indiaprisons} and US Prisons Data Set \cite{usprisons}. 
 Table \ref{table:independent} presents the scoring of the individual facets for these two data sets.  The US prisons data has an advantage over the India Prisons data sets due to availability of records at individual records, while Indian prisons data sets provides aggregate information. Using this method of breakdown analysis to arrive at a value, we see that it is easy to identify data sets that are more valuable relative to others.

 Our second example is in the context of an application; we consider an enterprise application of Job data normalization \cite{job} where descriptions of job requirements are normalized across several organizations. Data sets were used from two different sources, one from internal HR sources and another from a public repository \cite{pubjob}. In this scenario, the enterprise data set having higher score has more features relevant to the application. Data practitioners can easily use this tool to identify application relevant data sets from group of similar data sets.
 
 In summary, our approach provides a simple, yet feasible approach to assess  data sets both in the presence and absence of application context. In reality, data practitioners can provide relative importance of the facets ($\alpha$)  and the values  for facets and their scores can be obtained by closely studying domain related data sets.

\section{Related work}
\label{relatedwork}
Pricing of data and information has been studied in different forms since the start of the Internet economy, by practitioners from economics, management sciences and computer science~\cite{varian},\cite{pricinginformationgoods},\cite{querybasedpricing}, \cite{pricingfordatamarkets}, where practical issues in pricing of information goods is discussed, and theoretical pricing models have been proposed in different scenarios such as query-based pricing, a comparison of selling versus pay-per-use pricing and optimal pricing of datasets for unit and step pricing schemes for profit maximization using an economics framework. Further, specific work has also been done in developing pricing of personal data in different scenarios, as for instance in ~\cite{pricingprivatedata}. The term infonomics refers to the concept that information is an actual enterprise asset~\cite{douglaney}, and the work describes different conceptual ways to compute value of information.
In the context of data quality, ~\cite{adir} discuss content-based measurement methods for commonly used quality dimensions such as completeness, validity, accuracy and currency. In ~\cite{ballou} the authors study the various utilities of trading completeness over consistency in a given decision context and give guidance for various trade-offs. All these works provide data quality assessment in given contexts. 

The study of the various aspects that make up a data set has received significant interest recently with growing concerns about data safety and safeguards for sensitive data. Further, the adoption of machine learning techniques for various applications have brought into limelight issues of explanability and presence of bias in data models. There has been some work to address these issues through definition of datasheets ~\cite{datasheets}, where the authors suggest the creation of a data sheet with each data set, that provides details of the data set that are of interest in the context of fairness and transparency. Similarly, the concept of data nutrition label is introduced in ~\cite{datanutritionlabel}, for a similar goal. The SDoC or Supplier's Declaration of Conformity is discussed in ~\cite{sdoc} to provide transparent reporting services for trust in AI. 

In the context of categorical assessment of data value, through an enumeration of the various facets, we have not seen any earlier approach that studies this problem and we believe our work is the first to address this problem.

\section{Conclusions and Discussion}
Our work is a first step towards quantifying the value of a data set. Data usage is dependent on the several characteristics of the data and therefore we base our data value computation on an assessment of the different characteristics. We list the various facets that could potentially influence the data value and convert them to a  quantitative measure to enable data value scoring in an application-agnostic manner. The strength of our approach is that it is simple to get started, can be extended in a straightforward manner if additional facets need to be added, and improves its accuracy with more information and context.  
We also plan to expand this work by automating the extraction of relevant facets from the data sets and automating the computation of the scores, which will enable provision of data value as a service.
\label{conclusion}

\bibliographystyle{ACM-Reference-Format}

\onecolumn
\begin{scriptsize}
\begin{longtable}{| p{.2\textwidth} | p{.43\textwidth} |p{.04\textwidth} |p{.04\textwidth} |p{.04\textwidth} |p{.04\textwidth} |} 

   \caption{Scoring data values for publicly available data sets `Indian Prisons', `US prisons' (NS = Not Significant,T= Transactional, S= Survey, MM = Middle Management, TM = This Month, TY = This Year)}
     \label{table:independent}
     \\
    \hline
       \multirow{2}{*}{ Data Facet} & \multirow{2}{*}{Sub-Facet} & \multicolumn{2}{|c|}{India Prisons}  & \multicolumn{2}{|c|}{US Prisons}  \\
       
       & & Response & Score &Response &Score  \\
        \hline
     \multirow{1}{*}{Data Layout}
  & What is the data layout?  & Structured &1 & Structured &1 \\
 \hline
\multirow{2}{*}{Data Composition} &Are these instances of primary data types?  &Y &1 &Y &1\\
                 &Are all instances similar? & Y &1 &Y &1  \\   
 \hline
   \multirow{8}{*}{Data Format} & What is the format of the data set file? &CSV &1 & CSV &1  \\
                 &Does it have a schema? &Y &1 &Y &1  \\  
                 &Is it an export or query result of relational data? & N &0 &N &0 \\
                 &Does it have any standard? & N & 0  &N & 0  \\
                 &Is the data stored in proprietary formats. &N &1 &N &1 \\
                &Is it in normalized form?&Y &1&Y&1 \\
                
   \hline
     \multirow{7}{*}{Data quality} & Are all the fields complete? &Y &1 &N & 0 \\
     &Are there known missing instances?& N &1 &N &1  \\
     & Is the data accurate? & Y &1 & Y&1 \\
    & Is it known to have duplicates? &N &1 &N & 1 \\
    & Is the data Consistent within the data set?  &Y &1 &Y &1 \\
     &Does the data have Noise? &N &1 &N & 1  \\
    &Is it complete, with respect to the purpose it defines?&Y &1 &Y &1  \\
   & Does it fill the missing values in an existing data
set? & N &0 &N &0 \\
     &Is it error-free? & N &1 &N &1 \\
     \hline
     \multirow{2}{*}{Data Age} & How current is the data? & Recent & 0.75 &Recent &0.75 \\
     &Is there a known later instance of the data? & N & 1& N & 1 \\
     &How frequently does the data get outdated/updated? & Yearly &1 & Yearly &1\\ 
     \hline
     \multirow{2}{*}{Data Volume and Access} & What is the data size? & .5 GB &0.75 & .35 GB &0.5  \\
     & Is it expensive to store this data? & N &1 & N &1 \\
     \hline
     \multirow{2}{*}{Data Velocity} &How rapidly can it be said to be generated? & NS &1 & NS & 1  \\
    &Is it streaming data? & N &0 &N &0 \\
    \hline
   \multirow{8}{*}{Statistical Property} & Is
it suitable for classification models? &N&0  & Y &1 \\
     &Is it suitable for linear regression models?&Y &1& Y&1 \\
     &Is it suitable for clustering models?&Y&1 &Y &1 \\
&Has it been used in ML algorithms already? & Y &1 &Y&1\\
     &is  data uniformly distributed over different fields? & Y &0 &Y &0  \\
    &Was any sampling applied on the data to get this
sample?& N &1 & N &1\\
    & Is it time-series data? & N &0 & N&0  \\
    & Is it suitable for bivariate analysis? & Y & 1 &Y&1 \\
    &Is it suitable for multivariate analysis? & Y & 1 &Y&1 \\
     \hline
    \multirow{5}{*}{Data sensitivity} &Is it free of confidential information? & N &1 & Y&0 \\
    &Does it contain personal identifiable information? &N &0& Y&1 \\
     &Is it free of information to be retained for mandatory
purposes? &N &1 &N &1 \\
     &Does it have revenue or financial data? & N &0  &N &0\\
     & Does it have Medical Data/Health Data? & Y &1 & Y & 1 \\
     &Does it have protected attribute? & Y  & 0& Y &0 \\
     \hline
   
     \multirow{3}{*}{Data Sourcing} & How was the data obtained?  & T &0.5 & T &0.5 \\
     &Is this data easy for me to get, difficult for others?  & N &0 & N &0\\
    &Is the data aggregated from many sources or from
single source? & Single  &0 &Single& 1 \\
     &Is this publicly available? & Y &0 &Y &0 \\
    &Is this enterprise-generated? & N & 0 & N &0\\
    &Is this data machine generated? & N & 0 & N &0 \\
    &Are there known alternates for this data set? & N& 1 &N&1\\
    \hline
     \multirow{1}{*}{Data Updation} & Is the data updated Frequently?  & N & 1 &N&1\\
     \hline
     \multirow{1}{*}{Data Granularity} &Is it aggregate or summary information?  & Y &0 & N&1 \\
   \hline
      \multirow{2}{*}{Enterprise Aspects} &
      Which parts of the business process does it
contribute to? & HR &0 & HR &0  \\
     &At which hierarchy in the organization is the data
used? & MM &0.75 & MM & 0.75 \\
&Is the data already making money? & N &0 &N &0\\
   
  \hline 
  
  Frequency of use &When was the data last used? &
  TM &1 & TY &0.75 \\
   \hline
   \multirow{3}{*}{Data Transformation} &Is it known to have had data transformations? &
   N &0 & Y &1 \\
   & Is it encrypted data? & N &1 &N &1 \\
   &Is it anonymised data? & N  &0 & N &0 \\

 \hline
 \multirow{7}{*}{Legal and Access Aspects} &Is this data free of any legal restrictions in usage?&
 Y &1& Y&1 \\
 & Was this data acquired as part of some contract? & N &0 & N &0 \\
  &Are there any contractual obligations on the data? & N &1& N&1\\
  &If pertaining to 'information about people', was there consent to use? & N &0 & N&0\\
  &Are there legal restrictions on using this data? &N &0&N&0\\
  
  &Is it governed by some license? &N &0& N&0\\
  &Are there export restrictions? &N &1 &N&1\\
   & Is the data easy to access? ?& Y & 1 &Y &1 \\
 \hline
 \multirow{2}{*}{Data processing } &Are there tools or programs to cleanse the data? &
  N&1& N&1\\
  & Are there tools or programs to pre-process the data? & N &1 &N&1\\
  \hline
  \hline
  \multicolumn{2}{|c|}{Data Value}&\multicolumn{2}{|c|}{\textbf{44.25}} &\multicolumn{2}{|c|}{\textbf{49.25 }}\\
  \hline
  \hline

\end{longtable}
 \end{scriptsize}

\clearpage
\twocolumn

\clearpage
\onecolumn

\scriptsize
\begin{longtable}{| p{.15\textwidth} | p{.45\textwidth} |p{.08\textwidth} |p{.08\textwidth} |p{.08\textwidth} |p{.08\textwidth}|}
  \caption{Data sets of resumes from two different sources `Public Repository'
11
 and `Enterprise HR sources' for Data Value. Ones in Blue indicates questions relevant only in the context of application.}
    \label{table:dependent}
    \\
    \hline
    \hline
        \multirow{2}{*}{Data facet} & \multirow{2}{*}{Sub-facet} & \multicolumn{2}{|c|}{Public Repository} & \multicolumn{2}{|c|}{Enterprise HR Data} \\
        & & Response & score & Response & score\\
        \hline
        \multirow{1}{*}{Data Layout}
 & What is the data structure? & Unstructured & 0.25 & Structured &1 \\
 \hline
   \multirow{2}{*}{Data Composition} &Are these instances of primary data types?  &Y &1 &Y &1\\
                &Are all instances similar? &Y &1 &Y &1 \\   
\hline
    \multirow{10}{*}{Data Format} & What is the format of data set & txt &0& NA &0\\
                &Does it have a schema? & N &0 &Y  &1 \\  
                &Is it an export or query result of relational data? & N &0 &Y&1\\
                &Does it adhere to a standard?& N & 1 & Y & 1 \\
                &If in proprietary format, is it open? & N & 0  & Y & 1 \\
                &{Is it in normalized form?} & N &0 &N &0 \\
                
    \hline
    \multirow{9}{*}{Data quality} & Are all the fields complete? &Y &1&Y&1\\
    &Is it error-free?& N & 1&N &1\\
    & Are there known missing instances? & Y &0& N&1\\
    & Does it fill the missing values in an existing data
set? & Y &1&Y&1\\
    &Is it complete, with respect to the purpose it defines?  &Y &1 &Y&1\\
    &Is it known to have duplicates?&N &1 &N &1 \\
    &Does it complement or supplement an existing data
set?&Y  &1&Y &1\\
    &Is the data accurate? & Y&1 & Y &1\\
    &\textcolor{blue}{What is the precision?} & NA &0 & NA & 0 \\
    & \textcolor{blue}{What is the recall?} & NA  &0 & NA &0 \\
    &Is the data consistent within the data set? &Y &1&Y &1\\
    &Does the data have noise? & Y &0 & Y &0\\
    \hline
    \multirow{5}{*}{Data Age} & 
  \textcolor{blue}{Does the data become less useful with age?} & N & 1 & N & 1 \\
&\textcolor{blue}{Does the data gain value with age?} & N & 0 &N &0\\
&Is there a known later instance of the data? & Y  &0 & Y  & 0 \\
&How frequently does the data get outdated/updated? &Daily &0.25 &Daily &0.25\\
    \hline
    \multirow{1}{*}{Data Volume} & What is the size of the data? & 0.5GB & 0.5 & 1 GB &0.75\\
    \hline
    \multirow{2}{*}{Data Velocity} & How rapidly can it be said to be generated? & Very Fast & 0.5 & Very Fast &0.5 \\
    &Is it
streaming data? & N&0 & N  &0\\
    \hline
   \multirow{6}{*}{Statistical Property} & Is
it suitable for classification models? &N &0 & N&0 \\
    &Is it suitable for linear regression models? &N &0& N &0\\
    &Is it suitable for clustering models? &Y &1 &Y &1 \\
    &Has it been used in ML algorithms already? &Y &1 &Y &1\\
    &Was any sampling applied on the data to get this
sample? & Y &0& N&1 \\
    & Is it time-series data? & NA &0 & N &1\\
    &Is it suitable for bivariate analysis? &Y &1 & Y &1 \\
    &Is it suitable for multivariate analysis? &Y &1 & Y &1\\
    \hline
    \multirow{6}{*}{Data sensitivity} &Is it free of confidential information? & Y &0 & Y &0\\
    &Is it free of personal identifiable information? &Y &1 & Y & 1\\
    &Is it free of information to be retained for mandatory
purposes? &Y  &1 &N &1\\
    &Is it free of revenue or financial data? & Y &0 &Y &0 \\
    & Does it have Medical Data/Health Data? & N  &0 & N & 0\\
    &Does it have protective variables? &N &0 &N &0\\
    \hline
   
    \multirow{8}{*}{Data Sourcing } & How was the data obtained? & Webcrawler &0 & Invited/Survey &1 \\
    &Can this data be easily accessed by all? & Y &1& N &0\\
    &Is the data aggregated from many sources or from
single source? & S & 1 & S &1\\
    &Is this data easy for me to get, difficult for others? & N & 0 & N &0\\
    &Is this enterprise-generated? & N &0 & Y &1\\
    &Is this publicly available? & Y & 0 & N &1\\
    &Is this data machine generated? & N &1 & N &1\\
    &Are there known alternates for this data set? & N &1 & N &1\\
    \hline
    \multirow{1}{*}{Data Updation} & Is the data updated frequently? & N &1 & Y &0\\
    \hline
    \multirow{1}{*}{Data Usage} & \textcolor{blue}{How easy is it to utilise the data?} & Complex &0 &Complex & 0 \\
    
     \hline
     \multirow{1}{*}{Data Granularity} &Is it aggregate or summary information & N&0 & N&0\\
   
    \hline
     \multirow{7}{*}{Enterprise Aspects} &
    \textcolor{blue}{ Is the data already making money? }& N &0 &N  &1\\
     & \textcolor{blue}{Will it improve the efficiency of an existing
application or business process?} & Y & 1 & Y &1\\
& \textcolor{blue}{Does it introduce a new channel to reach to
customers?} & N &0 & N &0\\
& \textcolor{blue}{Does it complement an existing application?} & Y  &1 &  Y &1\\
     &\textcolor{blue}{Does it increase customer reach?} & N &0 &N&0\\
     &Which parts of the business process does it
contribute to? & HR & 1 & HR &1\\
&At which hierarchy in the organization is the data
used? & Others & 0.5 &Others & 0.5\\

   
 \hline 
   \multirow{1}{*}{Frequency of use} &When was the data last used? &
   last 5 yrs & 0.5 & last 5 yrs  &0\\
   & Is there a ’known’ future use? & N &0 &N &0\\
  \hline
  \multirow{2}{*}{Data Transformation} &Is it known to have had data transformations? &
  N &1& N&1\\
  & Is it encrypted data? & N &1&N&1\\
   &Is it anonymised data? & N&0 & Y&1\\

\hline
\multirow{7}{*}{Legal and Access Aspects} &Is this data free of any legal restrictions in usage? &
Y& 1 & N &0\\
& Was this data acquired as part of some contract? & N &1 &N &1\\
&Are there any contractual obligations on the data?& N &1 &N &1\\
 &If pertaining to ’information about people’, was there consent to use? & N &0& Y &1\\
 &Are there legal restrictions on using this data? & N &0 & N & 0\\
 
 &Is it governed by some license? &N &1 & N &1\\
 &Are there export restrictions? &N &1&N &1\\
 &Is the data easy to access? & Y & 1 & Y &1\\
\hline
 \multirow{2}{*}{Data Processing} & Are there tools or programs to cleanse the data? &
 Y& 1 &  Y &1\\
 & Are there tools or programs to pre-process the data? & N &0 &N &0\\
 \hline
 \multicolumn{2}{|c|}{Data Value} & \multicolumn{2}{|c|}{37}  & \multicolumn {2}{|c|}{45}\\
 \hline
\end{longtable}

\clearpage
\twocolumn


\end{document}